\documentclass[prd,preprint,nofootinbib]{revtex4}
\usepackage{framed}
\usepackage{bbold}
\usepackage{slashed}
\usepackage{graphicx,hyperref}
\usepackage{amsmath}
\usepackage{amssymb}
\usepackage{epsfig}
\usepackage{hhline}
\usepackage{soul}
\usepackage{tabularx,pifont,multirow}
\usepackage{enumitem}
\usepackage{colordvi}
\usepackage{soul}
\usepackage{xcolor}
\usepackage{yfonts}

\newcommand{\be}{\begin{equation}}
\newcommand{\ee}{\end{equation}}
\newcommand{\bea}{\begin{eqnarray}}
\newcommand{\eea}{\end{eqnarray}}

\definecolor{darkgreen}{rgb}{0,0.3,0.05}
\makeatletter                                                         %
\newcommand*\rel@kern[1]{\kern#1\dimexpr\macc@kerna}                  %
\newcommand*\widebar[1]{                                              %
  \begingroup                                                         %
  \def\mathaccent##1##2{                                              %
    \rel@kern{0.8}                                                    %
    \overline{\rel@kern{-0.8}\macc@nucleus\rel@kern{0.2}}             %
    \rel@kern{-0.2}                                                   %
  }                                                                   %
  \macc@depth\@ne                                                     %
  \let\math@bgroup\@empty \let\math@egroup\macc@set@skewchar          %
  \mathsurround\z@ \frozen@everymath{\mathgroup\macc@group\relax}     %
  \macc@set@skewchar\relax                                            %
  \let\mathaccentV\macc@nested@a                                      %
  \macc@nested@a\relax111{#1}                                         %
  \endgroup                                                           %
}                                                                     %
\makeatother                                                          %

\begin{document}

\preprint{\leftline{KCL-PH-TH/2021-{\bf 62}}}

\title{Geometrical origins of the Universe dark sector: string-inspired torsion and anomalies as seeds for inflation and dark matter} 

\vspace{0.2cm}

\author{{\bf Nick~E.~Mavromatos$^{\,a,\,b}$} }

\vspace{0.2cm}

\affiliation{$^a$Theoretical Particle Physics and Cosmology Group, Physics Department, King's College London, Strand, London WC2R 2LS, UK,
\vspace{0.1cm}\\
 $^{b}$ Physics Department, School of Applied Mathematical and Physical Sciences, National Technical University of Athens, 9, Heroon Polytechneiou Str., Zografou Campus, Athens 157 80, Greece.
}

\begin{abstract}
\vspace{0.05cm}
In a modest attempt to present potentially new paradigms in Cosmology, including its inflationary epoch, and initiate discussions, I  review in this article some novel, string-inspired cosmological models, which entail a purely geometrical origin of the Dark sector of the Universe but also of its observed matter-antimatter asymmetry. The models  contain gravitational (string-model independent, Kalb-Ramond (KR)) axion fields coupled to primordial gravitational anomalies via CP-violating interactions. The anomaly terms are four-space-time-dimensional remnants  of the Green-Schwarz counterterms appearing in the definition of the field strength of the spin-one antisymmetric tensor field of the (bosonic) massless gravitational string multiplet, which also plays the r\^ole of a totally antisymmetric component of torsion. I show how in such cosmologies the presence of primordial gravitational waves can lead to anomaly condensates and dynamical inflation of a "running-vacuum-model" type, without external inflatons, but also to leptogenesis in the radiation era due to anomaly-induced Lorentz and CPT Violating KR axion backgrounds. 
I also discuss how the torsion-related KR-axion could acquire a mass during the QCD epoch, thus playing the role of (a component of) Dark Matter. Phenomenological considerations of the inflationary and post-inflationary (in particular, modern) eras of the model are briefly discussed, including its potential for alleviating the observed tensions in the cosmological data of the current epoch.

\vspace{0.2cm}

\begin{flushleft}
\end{flushleft}
\end{abstract}

\maketitle


\tableofcontents

\section{Introduction and Summary: Phenomenological motivation}

The standard concordance model of Cosmology (also known as $\Lambda$CDM (cosmological-constant($\Lambda$)-Cold-Dark-Matter paradigm))~\cite{pdg} works very well in describing the large scale structure of the Universe, being in excellent agreement with a plethora of diverse cosmological observations, namely Supernovae SnIa, Cosmic Microwave Background (CMB), baryon acoustic oscillations (BAO) and gravitational lensing studies~\cite{Planck}. The data point towards the fact that the current-epoch 
energy density of the Universe  consists of 4.9\% baryonic
matter, 26.8\% dark (gravitating) matter and 68.3\%  a mysterious dark energy component, 
dominant at late eras (corresponding to redshifts $z\,<\,1$), which is held responsible for 
the accelerating expansion of the Universe, in a form compatible with the addition of 
a positive Cosmological Constant $\Lambda$ in Einstein's General Relativity (GR) theory (de Sitter space-time).\footnote{There is, though, the approach of \cite{subir}, in which the authors claim that the dark energy could be an artefact of interpretation and analyses of the supernovae data, which constitute the only direct evidence of the Universe acceleration so far, 
based on the assumption that we are idealised Friedman-Lemaitre-Robertson-Walker (FLRW) observers,
while the real Universe might be inhomogeneous and anisotropic at sufficiently large scales, to impact on the analyses. 
In our approach in this work we shall assume isotropy and homogeneity at cosmological distances, 
and build our cosmology around that.} On the other hand, several single-field models of inflation~\cite{inflation} have been falsified using CMB data~\cite{Planck}.  In this respect, the Starobinsky model of inflation~\cite{staro}, which is due to non-linear, higher-order-in-(scalar)-curvature corrections to GR, induced by quantum (conformal, ``trace") anomalies in effective field theories of (massless) matter in the early universe, seems to fit excellently the pertinent data~\cite{Planck}, provided one fixes the coefficient of the (anomaly-induced) higher-curvature corrections. A pressing question is how one can distinguish among the remaining models of inflation that have not been falsified by the data so far, and whether it would ever be possible to differentiate between models involving dynamical inflation, like Starobinsky's, and those based on fundamental inflaton fields. 

In this review, I will try to answer partially this question, by arguing that one needs to take into account the entirety of the data pertaining to the Universe evolution. I will suggest some novel directions towards dynamical inflation, different from Starobinsky's, which point towards a geometric origin of both inflation and Dark Matter (DM), and I will discuss briefly their phenomenology, which hopefully can lead to some (observable) distinguishing features from the currently available inflationary scenarios, including Starobinsky's. 

On another front, local (at cosmological scales) precision measurements of gravitational waves (GW) due to black hole (BH) mergers~\cite{ligo}, or studies of BH properties through imaging (``photographs'' )~\cite{eht1,eht2,eht3} and other techniques (e.g. pulsar timing~\cite{grt3}), have confirmed the validity of Einstein's theory of GR to a large extent, given that no deviations from it have been observed as yet~\cite{grt4,grt1,grt2}. Thus, it seems that both the global (cosmology) and local versions of GR provide good (classical) descriptions of the Universe.  

Nonetheless, recently there seem to be some persisting tensions in the cosmological data 
between early and late Universe~\cite{tensions}, which call for an explanation. One of them
(known as $H_0$ tension) points to a discrepancy between the value of the current-era Hubble parameter $H_0$ measured  in direct local observations 
($H_0 = 73.24 \pm 1.74 $~km/s/Mpc using 2016 Cepheids data~\cite{Htension},  
$H_0 = 73.48 \pm 1.66$~km/s/Mpc using 2018 Cepheids data~\cite{Htension1}
and $H_0 = 73.2 \pm 1.3 $~km/s/Mpc using more recent 2021 combined data~\cite{Htension2})
and that inferred from CMB measurements of the Planck Collaboration, based on $\Lambda$CDM fits ($H_0 = 67.27 \pm 0.60 $~km/s/Mpc in Planck 2018 data~\cite{Planck}). In addition, there are also some tensions between direct observations and simulations based on $\Lambda$CDM, which are associated with the growth of structure in the Universe (the so-called $\sigma_8$ or $S_8$ tension, where $\sigma_8$ denotes the root-mean-square (rms) fluctuations of the current-era matter density, within spheres of radius $8h^{-1}$, with $h=H_0/100$ the reduced Hubble parameter in the present epoch, and 
$S_8 = \sigma_8 \, \sqrt{\Omega_m/0.3}$, with $\Omega_m$ the current-era matter energy density in the Universe)~\cite{s8tension}.
Although it is still unclear whether these tensions point towards new physics, in the sense of departure from the $\Lambda$CDM paradigm, or admit more mundane astrophysical explanations~(see, e.g. \cite{efst,dust,dust2}), or even they will disappear in future data, being due to current statistical uncertainties, nonetheless their persisting nature so far prompted several theorists to seek models that can alleviate them~\cite{sol3,sol2,sol1}. 

One of the models that succeeds in alleviating simultaneously the $H_0$ and $\sigma_8$ tensions~\cite{solaepl}, 
which will be of interest to us in this article, is based on the so-called ``running vacuum model (RVM)" of cosmology~\cite{rvm,rvm2,rvm3,rvmfoss,solahere}. The latter is an effective cosmological framework in which the energy density of the vacuum is a function of even powers of the Hubble parameter $H(t)$ (as dictated by general covariance), and satisfies the equation of state of the de Sitter space time, despite its cosmic-time dependence (we note that a similar situation characterises the so-called non-critical string cosmology, with the cosmic time being identified with (the zero mode of) a string-world-sheet renormalization-group scale~\cite{emn}). 
The RVM describes a smooth evolution of the Universe  from inflation till the present era~\cite{rvmevol,rvmevol2,rvmthermo}, and is not characterised by initial singularities. In the RVM framework, inflation is 
due to the dominant (at early eras) non-linear $H^4$ terms in the cosmic vacuum energy density, and does not require external inflaton fields. The RVM is in agreement with the plethora of cosmological data~\cite{rvmsola}.  In the current era, the RVM energy density 
contains a cosmological constant term and a term proportional to $H^2$, the latter being responsible for observable deviations from the $\Lambda$CDM~\cite{rvmpheno1,rvmpheno,rvmpheno2,tsiapi} and potential tension alleviations~\cite{solaepl,solahere}. Moreover, it has been recently shown~\cite{rvmqft} that the RVM can be obtained as an effective theory of a (non-minimally coupled) quantum scalar field in a curved  cosmological space time.

On the front of GR, although all current observations agree excellently with it, nonetheless there are serious theoretical reasons to go beyond it, one of the most important of which is its unsuitability to provide a consistent, renormalizable theory of quantum gravity, due to the dimensionful nature of its coupling parameter, the Newton's constant G. 
One of the most successful, from a theoretical viewpoint, extensions of GR, which leads to a mathematically consistent formulation of quantum gravity, is string theory and its brane extensions~\cite{string}. Of course, string theory has its own drawbacks, the most important of which, in the opinion of the author of this review, is the lack of a unique ground state (landscape of string vacua). Nonetheless, string theory provides an elegant and 
mathematically consistent (gauge-invariant) framework in which (quantum) gravitational interactions are unified with the rest of the fundamental interactions in nature. This accounts for its enormous popularity over the past several decades.

In what follows we shall attempt to tackle the aforementioned  cosmological issues by considering some novel cosmologies based on a low-energy limit of strings. As we shall argue, in such an approach one can give a purely geometrical 
interpretation on both the dark-energy and dark-matter sectors of the Universe. Specifically, upon the assumption of a large-(cosmological)-scale homogeneity and isotropy of the Universe, we shall argue that a gravitational field theory model~\cite{ms1,ms2,basilakos,basilakos2} based on the degrees of freedom of the {\it massless} (bosonic) gravitational multiplet of the string~\cite{string}, 
provides an interesting extension 
of Cosmology, in the early epochs of which there are {\it gravitational anomalies}. 
We shall also link this string-inspired cosmology with the RVM framework~\cite{rvm,rvm2,rvm3,rvmfoss}, and in particular with a (phenomenological) variant of the RVM, called type II RVM~\cite{solaepl}, in which the energy density of the vacuum, in addition to being dependent on $H^{2n}$, $n \in \mathbb Z^+$, and containing terms involving $\dot H$, also contains an effective Newton's constant G$_{\rm eff}(t)$ which depends very mildly on the cosmic time $t$.  
We point out that the type II RVM succeeds in alleviating simultaneously the  $H_0$ and $\sigma_8$ tensions of the current data~\cite{solaepl,solahere}.

Within our string framework, we shall also argue~\cite{basilakos,basilakos2} that inflation is due to a condensation of primordial gravitational waves (GW), which, in the presence of gravitational anomalies at early eras, leads to de Sitter type contributions to the vacuum energy and inflation of RVM type~\cite{rvmevol,rvmevol2}, driven by an $H^4$ term in the vacuum energy density, without the need for external inflatons. At the exit from the RVM inflation in our string-inspired cosmology, chiral fermionic matter is generated, which is responsible for the cancellation of the primordial gravitational anomalies, leaving though chiral anomalies present during the post inflationary epochs. The latter are responsible for generating a non-perturbative mass (through instanton effects) for the KR axion during the QCD  epoch of the Universe. In this way, the KR axion can play the r\^ole of a DM component, and, if dominant, 
one obtains a geometric origin of DM, in view of KR-axion's connection to (a totally antisymmetric) torsion~\cite{torsion,kaloper,torsion2shap} in the string-inspired effective action.

An additional reason I would like to give for going beyond Einstein's GR theory and the standard concordance model of Cosmology is one which is usually attributed to the realm of particle physics, namely a microscopic explanation for the dominance of matter over antimatter in the Universe. According to the celebrated Sakharov's conditions~\cite{zakh}, 
in order to create baryon asymmetry and thus dominance of matter over antimatter in the Universe, it is necessary to have  violations of Charge conjugation symmetry (C), Charge-Parity (CP) and Baryon number (B)  in the early Universe, as well as departure from (thermal) equilibrium. The Standard Model (SM) of particle physics, at least in its hadron (quark) sector, where so far CP Violation (CPV) has been observed, although qualitatively satisfies these conditions, however it does not have~\cite{smbau,smbau2,smbau2b} sufficient amount of CPV to reproduce the observed Baryon Asymmetry in the Universe (BAU)~\cite{pdg}  
\begin{align}\label{bau}
\frac{n_b - n_{\overline b}}{n_b + n_{\overline b}} \simeq (8.4-8.9) \times 10^{-11} , 
\end{align}
for temperatures $T > 1$~GeV (where $n_{b(\overline b)}$ denotes the density of baryons (antibaryons) in the Universe).
This prompts the search for particle physics models with
considerable departure from the SM in their physical excitation spectra 
(supersymmetry, string-inspired  extra dimensional models {\it etc.}), which provide extra sources of CPV that could account for the observed BAU. 

I will argue in this review, however, that one may have a geometric origin of the BAU, due to CPT-Violating (CPTV) geometries in the early universe~\cite{sarkar1,sarkar2,sarkar3,sarkar4,sarkar5,sarkar6,sarkar7}, as a consequence of {\it torsion condensation}~\cite{basilakos,basilakos2,popl,popl2}, which imply that the observed BAU can be produced in minimal extensions of the SM spectrum, with right-handed neutrinos, via a leptogenesis stage~\cite{lepto}, which is succeeded by baryogenesis~\cite{baryolepto}.
Specifically, I will argue that the condensate of gravitational anomalies during the inflationary phase of the stringy RVM~\cite{basilakos,basilakos2}, induced by primordial GW,  implies Lorentz Violating (LV) backgrounds for the KR axion field, which remain undiluted at the exit from inflation and into the radiation era. Since such backgrounds admit a (gravitational) torsion interpretation, they couple universally to all fermions in the model~\cite{torsion,kaloper,torsion2shap}. 
In models containing (Majorana) right-handed neutrinos (RHN) in their spectra, the decay of the latter in the presence of the aforementioned KR-axion backgrounds implies lepton asymmetry, as a result of the asymmetric decay of the RHN onto SM leptons and antileptons~\cite{sarkar3,sarkar4,sarkar5,sarkar6,sarkar7}. The torsion interpretation of the LV KR axion background in string theory~\cite{string,kaloper}, implies then a {\it  geometric origin}  of the 
 cosmic {\it matter-antimatter asymmetry} in such models.
 
The structure of the review is as follows: In the next section \ref{sec:string}, we describe
the basic features of string-inspired low-energy effective field theories, with gravitational anomalies, that we shall make use of in our approach. 
In section \ref{sec:stringyrvm}, we describe the stringy RVM cosmological model and the crucial r\^oles played by the gravitational anomaly terms, primordial GW and the associated LV KR axion backgrounds in inducing an RVM inflation at early eras. A brief discussion on a potential origin of GW in a pre-RVM-inflationary era is also given. In section \ref{sec:postinfl} we discuss the post inflationary eras, which are free from
gravitational anomalies, and explain how the KR axion can become Dark Matter during the post-inflationary era, as a result of chiral anomalies, and also how it can lead to baryogenesis through LV and CPTV leptogenesis. 
Modern era phenomenology of the stringy RVM is briefly discussed in section \ref{sec:modern}, with the emphasis placed on the potentially observable deviations of the model from $\Lambda$CDM and potential alleviations of tensions in the current-era cosmological data. Finally, section \ref{sec:concl} contains our conclusions and outlook. 

\section{Elements of string-theory-inspired cosmologies with gravitational anomalies \label{sec:string}}

The (bosonic) {\it massless} gravitational multiplet of strings consists of~\cite{string}
a spin-0 (scalar)  dilaton, a symmetric spin-2 traceless tensor, $g_{\mu\nu}=g_{\nu\mu}$, which is identified as the graviton, and a spin-1 antisymmetric tensor (Kalb-Ramond (KR)) field
$B_{\mu\nu}=-B_{\nu\mu}$, where $\mu, \nu =0, \dots D-1$, with $0$ a temporal index, and $D$ the target-space dimension of the string. In our approach below, we shall be interested in strings compactified to four space-time dimensions, so from now on we take $D=4$. The world-sheet vertex operator of the field $B_{\mu\nu}$  (in the closed-string sector) is given by 
\begin{align}\label{bu1}
\int_{\Sigma^{(2)}} d^2 \sigma \, \varepsilon_{AB} \partial_A X^\mu \, \partial_B X^\nu~,
 \,\, \mu, \nu =0, \dots 3,  
\end{align}
where the integral is over the surface $\Sigma^{(2)}$, which corresponds to the string-tree-level world-sheet with the topology of a two-dimensional sphere $S^{(2)}$, of interest to our purposes here, given that string loop corrections, which would be associated with higher-genus world-sheet surfaces, are subdominant for weak string couplings we assume throughout; the indices $A,B=1,2$ are world-sheet indices, 
$\varepsilon_{AB}=-\varepsilon_{BA}$ is the world-sheet covariant Levi-Civita antisymmetric tensor, and $X^\mu$, $\mu=0, \dots 3$, are world-sheet fields, whose zero modes play the r\^ole of target-space coordinates. 
It can be seen straightforwardly (taking into account that the spherical-like surface $\Sigma^{(2)}$ has no boundary) that the operator \eqref{bu1} is invariant under the following U(1) gauge transformation in target space (which is not related to electromagnetism):
\begin{align}\label{bu2}
B_{\mu\nu} \, \to \, B_{\mu\nu} + \partial_\mu \Theta_\nu (X) - \partial_\nu \Theta_\mu (X), 
\quad \mu, \nu =0, \dots 3, 
 \end{align}
where $\Theta_\mu(X)$, $\mu=0, \dots 3,$ are  gauge parameters. This implies that the target-space effective action, which describes the low-energy limit of the string theory at hand, will be invariant under the U(1) gauge symmetry \eqref{bu2}, and, as such, it will depend only on the field strength 
of $B_{\mu\nu}$ :  $H_{\mu\nu\rho} = \partial_{[\mu} \, B_{\nu\rho]}$, where the symbol 
$[\dots ]$ indicates antisymmetrisation of the respective indices. However, in string theory, 
cancellation between gauge and gravitational anomalies in the extra dimensional space requires the introduction of Green-Schwarz counterterms~\cite{string}, which results in the modification of the field strength $H_{\mu\nu\rho}$  by the respective Chern-Simons (gravitational (``Lorentz'', L)  and gauge (Y)) anomalous terms (in form language, for notational convenience):
\begin{align}\label{GSH}
\mathbf{{\mathcal H}} &= \mathbf{d} \mathbf{B} + \frac{\alpha^\prime}{8\, \kappa} \, \Big(\Omega_{\rm 3L} - \Omega_{\rm 3Y}\Big),  \nonumber \\
\Omega_{\rm 3L} &= \omega^a_{\,\,c} \wedge \mathbf{d} \omega^c_{\,\,a}
+ \frac{2}{3}  \omega^a_{\,\,c} \wedge  \omega^c_{\,\,d} \wedge \omega^d_{\,\,a},
\quad \Omega_{\rm 3Y} = \mathbf{A} \wedge  \mathbf{d} \mathbf{A} + \mathbf{A} \wedge \mathbf{A} \wedge \mathbf{A},
\end{align}
where $\wedge$ denotes the exterior product among differential ($k,\ell$) forms (${\mathbf f}^{(k)} \wedge {\mathbf g}^{(\ell)} = (-1)^{k\, \ell}\, {\mathbf g}^{(\ell)} \wedge {\mathbf f}^{(k)}$). In the above expression, $\mathbf{A}$ denote the Yang-Mills gauge field one form, and $\omega^a_{\,\,b}$ is the spin connection one form, with the Latin indices $a,b,c,d$ being tangent space (SO(1,3)) indices. The Regge slope $\alpha^\prime=M_s^{-2}$, where 
$M_s$ is the string mass scale, which is in general different from the reduced Planck scale 
in four space-time dimensions that enters the definition of the four-dimensional gravitational
constant $\kappa = \sqrt{8\pi\, {\rm G}} = M_{\rm Pl}^{-1}$, with $M_{\rm Pl} =  2.43 \times 10^{18}$~GeV  (we work in units of $\hbar=c=1$ throughout).

To lowest order in $\alpha^\prime$, {\it i.e.} to quadratic order in a derivative expansion, the low-energy effective four-dimensional action corresponding to the bosonic massless string multiplet, reads~\cite{string} (for notation and conventions, see \cite{basilakos2}):
\begin{align}\label{sea}
S_B  =\; \int d^{4}x\sqrt{-g}\Big( \dfrac{1}{2\kappa^{2}} [-R + 2\, \partial_{\mu}\Phi\, \partial^{\mu}\Phi] - \frac{1}{6}\, e^{-4\Phi}\, {\mathcal H}_{\lambda\mu\nu}{\mathcal H}^{\lambda\mu\nu} + \dots \Big),
\end{align}
with the ellipses $\dots$ denoting higher-derivative terms, and possible dilaton potentials (arising from string loops or other mechanisms in effective string-inspired models, such as 
non-critical-string cosmologies~\cite{aben,emn}, and pre-Big-Bang scenarios~\cite{prebb}.). In our approach we shall consider the dilaton field as fixed to an appropriate constant value, corresponding to minimisation of its potential, 
so that the string coupling $g_s=\exp (\Phi)$ is fixed to phenomenologically acceptable values~\cite{string}. The torsion~\cite{torsion} interpretation of $\mathcal H_{\mu\nu\rho}$ arises by noticing that one can combine the quadratic in $\mathcal H_{\mu\nu\rho}$  terms of \eqref{sea} with the Einstein-Hilbert curvature scalar term $R$ in a generalised curvature scalar $\overline R(\overline \Gamma)$ with respect to a generalised connection:
${\overline \Gamma}_{\mu\nu}^{\rho} = \Gamma_{\mu\nu}^\rho + \frac{\kappa}{\sqrt{3}}\, {\mathcal H}_{\mu\nu}^\rho  \ne {\overline \Gamma}_{\nu\mu}^{\rho}$,
where $\Gamma_{\mu\nu}^\rho = \Gamma_{\nu\mu}^\rho$ is the torsion-free Christoffel symbol. Since the KR field strength satisfies $\mathcal H^\mu_{\nu\rho} = -
\mathcal H^\mu_{\rho\nu}$, it plays the r\^ole of contorsion. This contorted geometry 
contains only a totally antisymmetric component of torsion~\cite{torsion}.\footnote{Using local field redefinition ambiguities~\cite{string,kaloper,amb1,amb2,amb3,amb4}, which do not affect the perturbative string scattering amplitudes~\cite{equiv1,equiv2}, one can extend the torsion interpretation of $\mathcal H$ to 
${\mathcal O}(\alpha^\prime)$ effective actions, which include fourth-order derivative terms.} This is a distinguishing feature of the string model from other generic torsion cosmologies (e.g. \cite{tors1,tors2}), in which the torsion has more components.
Moreover, as the string multiplet necessarily contains a graviton field, this string-inspired gravitational theory is different from teleparallel gravity and cosmology~\cite{tele}, where torsion mimics the gravitational field.

The modification (\ref{GSH})
leads to the Bianchi identity (in differential form language)~\cite{string}
\begin{equation}\label{modbianchi}
\mathbf{d} \mathbf{{\mathcal H}} = \frac{\alpha^\prime}{8 \, \kappa} {\rm Tr} \Big(\mathbf{R} \wedge \mathbf{R} - \mathbf{F} \wedge \mathbf{F}\Big)
\end{equation}
where $\mathbf{F} = \mathbf{d} \mathbf{A} + \mathbf{A} \wedge  \mathbf{A}$ is the Yang-Mills field strength two form  and $\mathbf{R}^a_{\,\,b} = \mathbf{d} \omega^a_{\,\,b} + \omega^a_{\,\,c} \wedge \omega^c_{\,\,b}$ the curvature two form, and the trace (Tr) is over gauge and Lorentz group indices respectively.
The non zero quantity on the right hand side  of \eqref{modbianchi} is the ``mixed (gauge and gravitational) quantum anomaly''~\cite{alvarez}. 
In the (more familiar) component form, the identity \eqref{modbianchi}, becomes:
\begin{align}\label{modbianchi2}
 \varepsilon_{abc}^{\;\;\;\;\;\;\;\;\;\mu}\, {\mathcal H}^{abc}_{\;\;\;\;\;\; ;\mu}
 =  \frac{\alpha^\prime}{32\, \kappa} \, \sqrt{-g}\, \Big(R_{\mu\nu\rho\sigma}\, \widetilde R^{\mu\nu\rho\sigma} -
F_{\mu\nu}\, \widetilde F^{\mu\nu}\Big) \equiv \sqrt{-g}\, {\mathcal G}(\omega, \mathbf{A}),
\end{align}
where the semicolon denotes covariant derivative with respect to the standard
Christoffel connection, and
$ \varepsilon_{\mu\nu\rho\sigma} = \sqrt{-g}\,  \epsilon_{\mu\nu\rho\sigma}, \quad \varepsilon^{\mu\nu\rho\sigma} =\frac{{\rm sgn}(g)}{\sqrt{-g}}\,  \epsilon^{\mu\nu\rho\sigma}$, denote the gravitationally covariant Levi-Civita tensor densities, totally antisymmetric in their indices, with $\epsilon_{\mu\nu\rho\sigma}$ ($\epsilon_{0123} = +1$, {\emph etc.}) the Minkowski-space-time Levi-Civita totally antisymmetric symbol. 
The symbol
$\widetilde{(\dots)}$
over the curvature or gauge field strength tensors denotes the corresponding duals, defined as $\widetilde R_{\mu\nu\rho\sigma} = \frac{1}{2} \varepsilon_{\mu\nu\lambda\pi} R_{\,\,\,\,\,\,\,\rho\sigma}^{\lambda\pi}$ and  $\widetilde F_{\mu\nu} = \frac{1}{2} \varepsilon_{\mu\nu\rho\sigma}\, F^{\rho\sigma}$, respectively.
The mixed-anomaly term is a total derivative
\begin{align}\label{pontryaginA}
&\sqrt{-g} \, \Big(R_{\mu\nu\rho\sigma}\, \widetilde R^{\mu\nu\rho\sigma} - F_{\mu\nu}\, \widetilde F^{\mu\nu} \Big) = \sqrt{-g} \, {\mathcal K}^\mu (\omega, \mathbf A)_{;\mu} = \partial_\mu \Big(\sqrt{-g} \, {\mathcal K}^\mu (\omega, \mathbf A) \Big) \nonumber \\
&= 2 \, \partial_\mu \Big[\epsilon^{\mu\nu\alpha\beta}\, \omega_\nu^{ab}\, \Big(\partial_\alpha \, \omega_{\beta ab} + \frac{2}{3}\, \omega_{\alpha a}^{\,\,\,\,\,\,\,c}\, \omega_{\beta cb}\Big) - 2 \epsilon^{\mu\nu\alpha\beta}\, \Big(A^i_\nu\, \partial_\alpha A_\beta^i + \frac{2}{3} \, f^{ijk} \, A_\nu^i\, A_\alpha^j \, A_\beta^k \Big)\Big],
\end{align}
where $i,j,k$ denote gauge group indices, with $f^{ijk}$ the gauge group structure constants.

In our
four dimensional cosmology~\cite{basilakos,basilakos2,ms1,ms2} we shall {\it not} cancel the anomalies. 
In fact, we shall assume that only fields of the bosonic degrees of freedom of the massless gravitational string multiplet appear as external fields in the effective action describing the dynamics of the early Universe. Chiral fermionic and gauge matter are generated at the end of the inflationary period as we shall discuss later on, in section \ref{sec:postinfl}. 
With the above assumptions, one may implement the Bianchi identity \eqref{modbianchi} as a {\it constraint} in a path-integral, via a pseudoscalar (axion-like) Lagrange multiplier field $b(x)$. After the ${\mathcal H}_{\mu\nu\rho}$ path-integration, then, one arrives at an effective action for the dynamics of the early epoch of the string-inspired Universe~\cite{basilakos,basilakos2,ms2}, which, upon the assumption of constant dilatons, contains only gravitons and the now dynamical field $b(x)$, canonically normalised, without potential, which corresponds to the massless string-model-independent gravitational (or KR)  axion field~\cite{kaloper,svrcek}: 
\begin{align}\label{sea4}
S^{\rm eff}_B =
\; \int d^{4}x\, \sqrt{-g}\Big[ -\dfrac{1}{2\kappa^{2}}\, R + \frac{1}{2}\, \partial_\mu b \, \partial^\mu b  -
 \sqrt{\frac{2}{3}}\,
\frac{\alpha^\prime}{96 \, \kappa} \, {\mathcal K}^\mu (\omega)\, \partial_\mu b(x)   + \dots \Big],
\end{align}
where we have used \eqref{pontryaginA}, setting $\mathbf A=0$, and performed appropriately the integration by parts, taking into account that fields and their first derivatives vanish at space-time infinity. We note that classically, in (3+1) dimensional space-times,  the duality between $\mathcal H_{\mu\nu\rho}$ and $b(x)$ is provided by the relation (corresponding to saddle points of the $\mathcal H$ path-integral)~\cite{kaloper,aben}
\begin{align}\label{dual}
-3\sqrt{2} \, \partial_\sigma b = \sqrt{-g} \, \epsilon_{\mu\nu\rho\sigma} \, \mathcal H^{\mu\nu\rho}.
\end{align}
The ellipses $\dots$ in \eqref{sea4} denote subdominant, for our purposes, higher derivative terms (in fact an infinity of them), but also 
other axions, arising from compactification in string theory~\cite{svrcek}, which 
have been discussed in \cite{ms2}, but will not be the focus of our present study. 
The reader should notice the presence of anomalous CP-violating couplings of the KR axion to gravitational anomalies in the action \eqref{sea4}. These will play an important role in inducing an RVM inflation in our string-inspired cosmology~\cite{basilakos,basilakos2,ms1,ms2}.

We note at this stage, that had we kept gauge fields in our early-universe cosmology as external fields, the KR axion field would also exhibit lagrangian couplings of the form
$\propto b(x) {\rm Tr} \Big(\mathbf F_{\mu\nu} \, \widetilde{\mathbf F}^{\mu\nu}\Big)$. Such terms would not contribute to the stress tensor, being topological. This needs to be contrasted with the 
gravitational anomaly terms in \eqref{sea4}, whose variation with respect to the metric field $g_{\mu\nu}$ yields non-trivial results:
$\delta \Big[ \int d^4x \sqrt{-g} \, b \, R_{\mu\nu\rho\sigma}\, \widetilde R^{\mu\nu\rho\sigma} \Big] = 4 \int d^4x \sqrt{-g} \, {\mathcal C}^{\mu\nu}\, \delta g_{\mu\nu} = -
4 \int d^4x \sqrt{-g} \, {\mathcal C}_{\mu\nu}\, \delta g^{\mu\nu}$, where ${\mathcal C}^{\mu\nu} \equiv  -\frac{1}{2}\, \Big[v_\sigma \, \Big( \varepsilon^{\sigma\mu\alpha\beta} R^\nu_{\, \, \beta;\alpha} +
\varepsilon^{\sigma\nu\alpha\beta} R^\mu_{\, \, \beta;\alpha}\Big) + v_{\sigma\tau} \, \Big(\widetilde R^{\tau\mu\sigma\nu} +
\widetilde R^{\tau\nu\sigma\mu} \Big)\Big]$, is the (tracelss) Cotton tensor~\cite{jackiw} with $v_{\sigma} \equiv \partial_\sigma b = b_{;\sigma}, \,\,v_{\sigma\tau} \equiv  v_{\tau; \sigma} = b_{;\tau;\sigma}$.
Taking into account conservation properties of the Cotton tensor\cite{jackiw}, 
${\mathcal C}^{\mu\nu}_{\,\,\,\,\,\,\,;\mu} = \frac{1}{8}\, v^\nu \, R^{\alpha\beta\gamma\delta} \, \widetilde R_{\alpha\beta\gamma\delta}$, we observe that the Einstein's equations stemming from \eqref{sea4} read: 
\begin{align}\label{einsteincs}
R^{\mu\nu} - \frac{1}{2}\, g^{\mu\nu} \, R  - \,  {\mathcal C}^{\mu\nu} = \kappa^2 \, T^{\mu\nu}_{\rm matter},
\end{align}
where $T^{\mu\nu}_{\rm matter}$ denotes a matter stress tensor, which in our cosmology 
includes only the KR axion-like fields (in more  general situations  $T^{\mu\nu}_{\rm matter}$ does {\it not} contain couplings to curvature or derivatives of the metric tensor). From the properties of the Einstein and Cotton tensors, stated above, we observe that the matter stress tensor is not conserved but it satisfies: 
$T^{\mu\nu}_{\rm matter \, ; \mu} -  {\mathcal C}^{\mu\nu}_{\,\,\,\,\,\,\,;\mu} =0$. 
The presence of the Cotton tensor in this conservation equation  indicates exchange of energy between the KR axion field and the gravitational anomaly term, in a way consistent with diffeomorphism invariance and general covariance~\cite{basilakos}. For FLRW geometries, the gravitational anomaly terms vanish, but this is not the case for fluctuations about the FLRW background which violate CP invariance, for instance gravitational-wave (GW) perturbations, as we discuss in the next section \ref{sec:stringyrvm}.

\section{String-inspired running vacuum model (stringy RVM)\label{sec:stringyrvm} } 

In this section we proceed to demonstrate how the stringy cosmological model of the previous section leads to an inflationary RVM cosmology in the presence of primordial GW. To this end, we first discuss a potential origin of GW in a pre-RVM inflationary epoch, in which, as we shall argue below, stiff axion matter is present.

\subsection{Stiff axion matter and origin of primordial gravitational waves \label{sec:preinfl}}

In the absence of CP violating (e.g. GW) perturbations of the FLRW metric, the anomalous couplings of the KR axion with the gravitational Chern-Simons terms in the action \eqref{sea4} are absent. If one assumes that such a gravitational action describes the space-time dynamics in the early Universe, then, in view of the fact that the KR axion in the theory has no potential, as a consequence of its implementation as a Lagrange multiplier of the Bianchi constraint \eqref{modbianchi}, one observes that the KR axion behaves as a {\it stiff} matter in this cosmological setting, 
i.e. a fluid with equation of state $w=+1$, assuming homogeneity and isotropy in the Universe. We note at this point that stiff matter dominance was assumed to characterise the 
early Universe also in the model of ref.~\cite{stiff1}, but in that model stiff matter consisted of baryons and not axions as in our  case. From a generic point of view, stiff matter eras have also been considered in cosmology in the work of ref.~\cite{stiff2}, but from a rather different perspective than ours. 

\begin{figure}[!h]
\centering\includegraphics[width=4.5in]{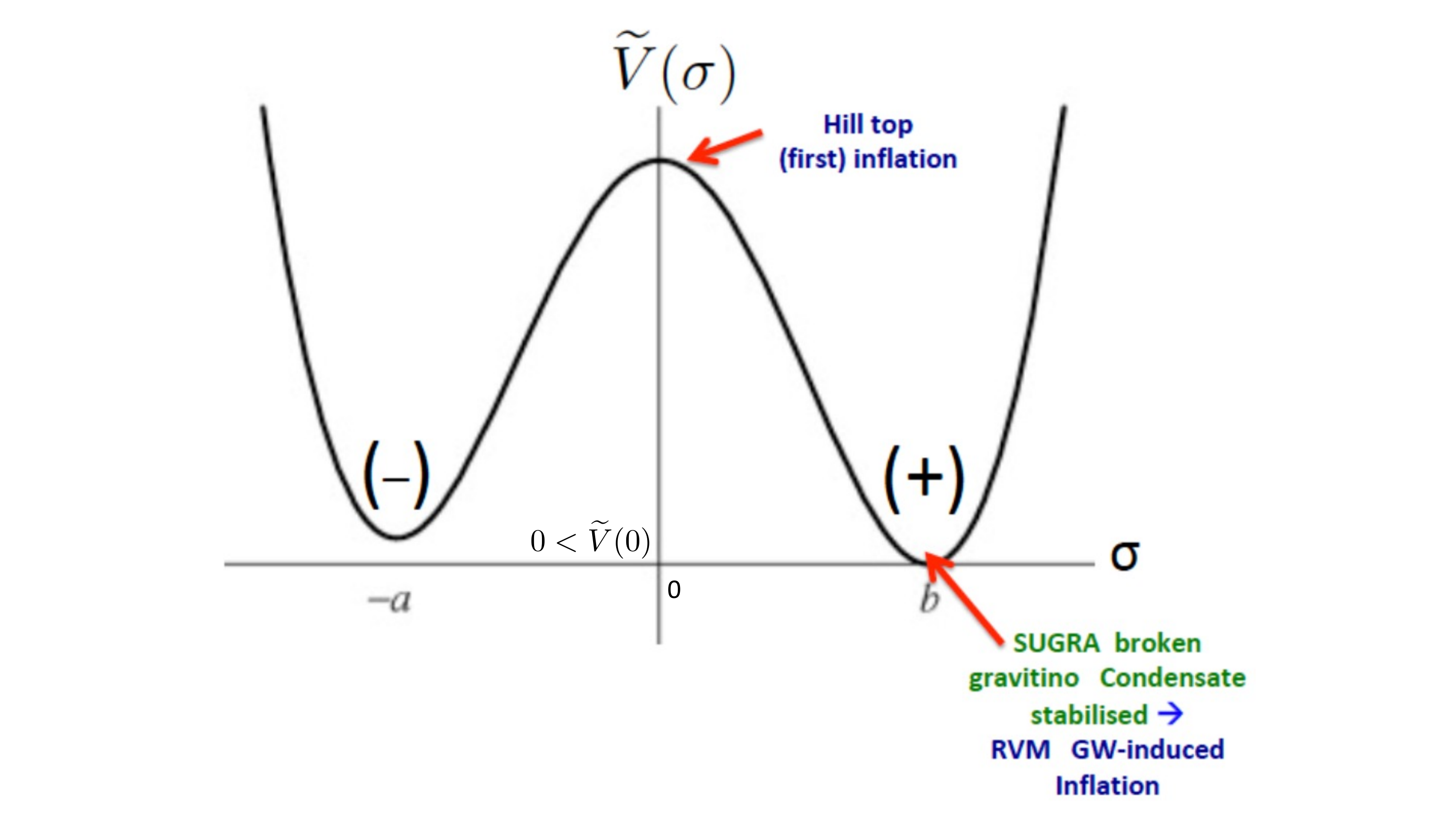}
\caption{The double well gravitino potential for dynamical supergravity breaking~\cite{brokensugra,ms2}, in which there is a statistical in origin bias between its two minima, due to their different occupation probabilities, as a result of percolation effects~\cite{ovrut1,ovrut2}. This can lead to the creation of unstable domain walls, whose non-spherically-symmetric annihilation will lead to GW.}
\label{fig_sugra}
\end{figure}

If such a phase characterises the early Universe, then before considering the 
effects of GW perturbations of the FLRW metric on the anomalies, we should examine the physical conditions which would lead to primordial GW in first place.
Various scenarios for this have been discussed in \cite{ms2}. Here we shall concentrate only on one of them, which is in accordance with the basic assumption of our approach~\cite{basilakos}, namely that the early universe effective action contains only fields from the massless gravitational multiplet of strings. In the case of superstrings, one may embed the model \eqref{sea4} into an appropriate supergravity model, in which the supersymmetric partner of gravitons, the gravitino $\psi^\mu$ (which also belongs to the gravitational supermultiplet of strings), can condense, leading to dynamically broken supergravity and a massive gravitino~\cite{brokensugra}. We assume~\cite{ms1,ms2}, then, that, after the Big Bang in this cosmology, some supergravity action is in operation with dynamically broken supergravity via a gravitino (scalar) condensate $\sigma = \langle \overline \psi_\mu \, \psi^\mu\rangle$, e.g. via the mechanism of \cite{brokensugra}. Depending on the parameters of the theory, the gravitino can become very massive, with a mass near the Planck scale~\cite{brokensugra}, and as such it can be integrated out in the effective action, leaving us with the light degrees of freedom of the massless gravitational multiplet of strings. The effective potential $\widetilde V(\sigma) $ has been calculated in
the concrete (and quite representative for our purporses here) case of $N=1$ four dimensional supergravity, and is found to have a double-well shape, with two
vacua, $\pm$, at which it assumes local minimum positive values, in agreement with the breaking of supergravity (and, hence, global supersymmetry as well)~\cite{brokensugra} (see fig.~\ref{fig_sugra}). 
Percolation effects can lead to different occupation numbers of the two ($\pm$) vacua 
({\it cf.} fig.~\ref{fig_sugra}), so that the $Z_2$ discrete symmetry is (statistically) broken\cite{ovrut1,ovrut2}. Such a bias leads to the formation of unstable domain walls, whose non-spherically-symmetric collapse (or collisions) leads to the formation of primordial GW~\cite{ms2,dw1,dw2}. 

\begin{figure}[!h]
\centering\includegraphics[width=4.5in]{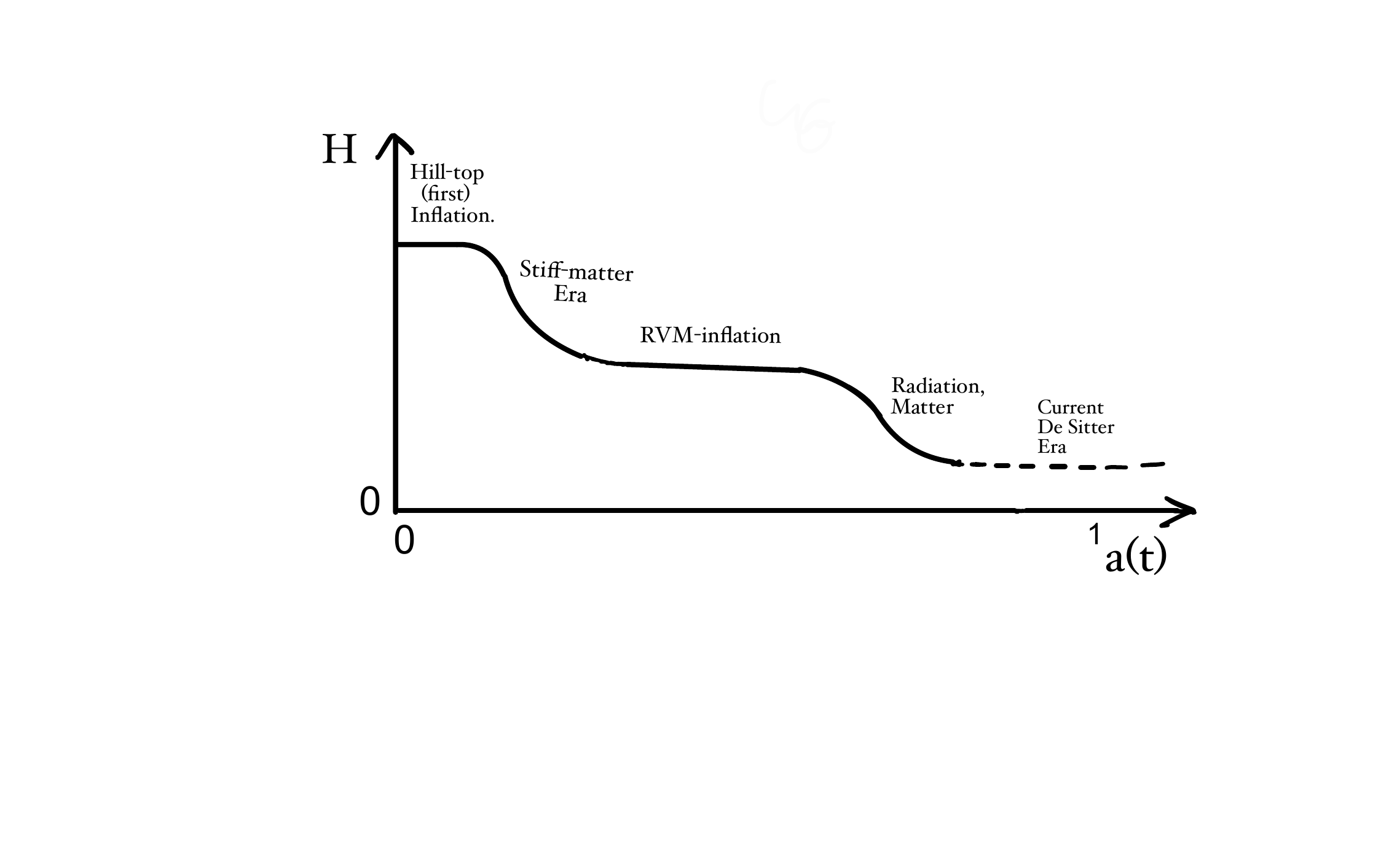}
\vspace{-2cm}
\caption{Diagram of the Hublle parameter ($H$) vs. the scale factor a(t), demonstrating the evolution of the stringy RVM universe~\cite{ms1}, from the Big Bang (a=0) till the current era (a=1). There are two inflationary epochs, separated by a stiff-axion-matter dominated era. The first (hill-top) inflation, near the Big Bang, occurs in models with dynamical supergravity breaking, as a result of gravitino condensate fields, which can play the r\^ole of the intlaton field. Its duraton depends on parameters of the supergravitry theory. The main inflation, which has observable consequences, succeeds the stiff-axion-matter era, and is due to GW condensates that lead to a RVM-type inflation, which is not characterised by external inflaton fields, but is due to the non-linear dependence of the RVM energy density on the quartic power of the Hubble parameter.}
\label{fig_rvminfl}
\end{figure}

Moreover, in such scenarios, one may have a hill-top first inflation~\cite{ellisinfl}, near the origin, where the gravitino condensate field, which plays the r\^ole of the inflaton, takes on small values, near $\sigma=0$ ({\it cf.} fig.~\ref{fig_sugra}). This inflation is not necessarily slow roll, and has no observable consequences. This first inflationary era is responsible for the washing out of any spatial inhomogeneities and anisotropies at large scales, during the stiff-KR-axion-matter era, which succeeds this phase. 
One can show~\cite{ms1} that the effective action during this inflation has imaginary parts, whose magnitude is such that the life time of the first inflationary phase is of order $1/H^{\rm first}_{\rm infl}$, where $H^{\rm first}_{\rm infl} \simeq {\rm constant}$ is the corresponding Hubble parameter. 
At the end of the first inflation, the gravitino condensate settles in the lowest of the two vacua ((+) vacuum in fig.~\ref{fig_sugra}), where the supergravity is dynamically broken  and the gravitino field is massive, and thus is integrated out of the effective field theory. The latter then consists only of massless gravitons and KR axions (under the assumption of constant dilatons). The exit from this first inflation is characterised~\cite{ms1,ms2} by stiff-KR-axion-``matter" dominance, and interpolates between the first inflation and the GW-induced RVM inflation ({\it cf.} fig.~\ref{fig_rvminfl}), which arises as a result of the condensation of the primordial GW that were formed due to the collapse of the domain walls. This will be discussed in the next subsection.

\subsection{GW condensation and RVM-type Inflation \label{rvminfl}} 

To demonstrate the existence of an RVM type inflation as a consequence of primordial GW condensates~\cite{basilakos,basilakos2}, we first assume the existence of an inflationary phase, with (approximately) constant Hubble parameter $H_I$, and compute the anomaly condensate, integrating over graviton fluctuations of GW CP-Violating (CPV) type. Then, we demonstrate~\cite{ms2} that an RVM equation of state and an appropriate RVM form~\cite{rvm,rvm2,rvmfoss} for the Universe vacuum energy density arise. In this way, inflation is obtained in the early stages of the RVM evolution~\cite{rvmevol,rvmevol2}, due to the dominant $H_I^4$ terms in the energy density, whose presence is the exclusive result of the condensate of gravitational anomalies~\cite{basilakos,basilakos2}. Below we derive these results in some detail.

To this end, let us first compute the gravitational anomaly condensate in the presence of GW perturbations, which we assume to be of the form: 
 \begin{align}\label{metric2}
 ds^2 = dt^2 - a^2(t) \Big[(1 - h_+(t,z))\, dx^2 + (1 + h_+(t,z))\, dy^2 + 2h_\times (t,z)\, dx\, dy + dz^2 \Big],
 \end{align}
 in the usual notation for the polarisation of the gravitational waves. 
 Assuming an approximately constant Hubble parameter $H \simeq$ constant, one may 
 integrate over GW perturbations (graviton modes), with spatial momenta of magnitude $k$  up to an Ultra-Violet (UV)  cutoff $\mu$, to obtain the anomaly condensate~\cite{stephon,basilakos2}:
 \begin{align}\label{rrt2}
  \langle R_{\mu\nu\rho\sigma}\, \widetilde R^{\mu\nu\rho\sigma} \rangle  &\simeq  
  \frac{16}{a^4} \, \kappa^2\int^\mu \frac{d^3k}{(2\pi)^3} \, \frac{H^2}{2\, k^3} \, k^4 \, \Theta  = \frac{1}{\pi^2} \Big(\frac{H}{M_{\rm Pl}}\Big)^2 \, \mu^4\, \Theta   \nonumber \\
 &= \frac{2}{3\pi^2} \frac{1}{96 \times 12} \,  \Big(\frac{H}{M_{\rm Pl}}\Big)^3 \, \Big(\frac{\mu}{M_{\rm Pl}}\Big)^4 \,  M_{\rm Pl}\, \times \, \, {\mathcal K}^0 (t)\,,
\end{align}
to leading order in the slow-roll parameter 
\begin{align}\label{theta}
 \Theta = \sqrt{\frac{2}{3}}\, \frac{\alpha^\prime \, \kappa}{12} \, H \,  {\dot {\overline b}} \, \ll \, 1~,
  \end{align}
with the overdot denoting derivative with respect to the cocmic time $t$. It should be stressed that this is a (weak) quantum gravity computation, about an FLRW de Sitter background. The effective gravitational action is given by \eqref{sea4}. In the absence of GW CPV metric fluctuations, the gravitational anomalies vanish. So the result \eqref{rrt2} is a result exclusive of situations where GW are present~\cite{stephon,basilakos,basilakos2}. Assuming isotropy and homogeneity, which in our (broken supergravity) scenario, discussed in the previous subsection \ref{sec:stringyrvm}\ref{sec:preinfl},  are guaranteed 
by the presence of the first hill-top inflation ({\it cf.} fig.~\ref{fig_rvminfl}), the equation of motion for the KR axion field $b(x)$, stemming from \eqref{sea4}, in the presence of the (temporal component of the) anomaly current $\mathcal K^0$, 
admits the solution~\cite{basilakos,basilakos2}
\begin{align}\label{krbeom2}
\dot{\overline{b}}  =  \sqrt{\frac{2}{3}}\, \frac{\alpha^\prime}{96 \, \kappa} \, {\mathcal K}^{0}.
\end{align}
Moreover, from the anomaly equation \eqref{pontryaginA} (ignoring the gauge part), using \eqref{rrt2}, we can calculate the anomaly current at a cosmic time $t$ since the onset of the RVM inflationary era at $t=0$\cite{basilakos}:
\begin{eqnarray}\label{k02}
{\mathcal K}^0 (t) = {\mathcal K}^0_{\rm begin} (t=0) \, \exp\Big[  - 3H\, t \, \Big( 1  -  
\frac{1}{3\,\pi^2 \times 18 \times  96}\, \Big(\frac{H}{M_{\rm Pl}}\Big)^2 \, \Big(\frac{\mu}{M_s}\Big)^4 \Big)\Big],
\end{eqnarray}
We then observe that, on setting the UV cutoff $\mu$ such that $\mu/M_s \simeq 15 \, \Big(M_{\rm Pl}/H\Big)^{1/2}$, we may ensure the (approximate) contstancy of $\mathcal K^0$ for the entire duration of the inflationary era:
\begin{align}\label{k0const}
\mathcal K^0 (t) = {\rm constant} , \quad t \in (0, t_{\rm end}),
\end{align}
 Taking into account the Planck-Collaboration-data result for the inflationary scale~\cite{Planck} 
 \begin{align}\label{PlHI}
 \frac{H_I}{M_{\rm Pl}} \simeq 10^{-5},
 \end{align}
 and identifying $H$ in \eqref{k02} with $H_I$, $H=H_I$, we observe that \eqref{k0const} requires $\mu \simeq \mathcal O(10^3\,M_s )$. On combining the slow-roll condition of $b(t)$ \eqref{theta} with the assumption of subplanckian values of $\mu$, so that no transplanckian graviton modes exist, and thus our model respects the transplanckian conjecture, it can be shown~\cite{ms1} that the string mass scale can be restricted to the range 
 \begin{align}\label{msscale}
 2.6 \times 10^{-5}\, M_{\rm Pl}  \lesssim M_s \lesssim 10^{-4}\, M_{\rm Pl}~.
 \end{align} 

Under \eqref{k0const}, one then obtains from \eqref{krbeom2} a solution for $\dot b$ which is
consistent with the Planck Collaboration slow-roll data~\cite{Planck}, provided we set~\cite{basilakos,basilakos2} 
\begin{align}\label{slowrollb} 
\dot b = \sqrt{2\epsilon} H  M_{\rm Pl}, \quad \epsilon = \mathcal O(10^{-2}), \quad H \simeq {\rm constant}, 
\end{align}
which implies 
\begin{align}\label{slowrollbint}  
b(t) = \overline b(0) + \sqrt{2\epsilon} H  M_{\rm Pl} \, t, \quad H \simeq {\rm constant}, 
\end{align}
with $\overline b(0)$ a boundary condition for the field $b(x)$ at the onset of inflation.
This is a spontaneously Lorentz-Violating (LV) solution ({\it cf.} also \eqref{dual}). It implies that $\dot b$ remains undiluted at the end of inflation and well onto the radiation era~\cite{basilakos,basilakos2,ms1,ms2}, where it induces LV and CPT-Violating (CPTV) leptogenesis in models that involve right-handed neutrinos (RHN)~\cite{sarkar3,sarkar4,sarkar5,sarkar6,sarkar7} (see section~\ref{sec:postinfl}).

The condensate \eqref{rrt2} also leads to the  condensate $\langle b R_{\mu\nu\rho\sigma} \, \widetilde R^{\mu\nu\rho\sigma} \rangle$, which, upon considering:\footnote{We note that the condition 
\eqref{b0} does not violate the transplanckian conjecture, because the effective gravitational action \eqref{sea4} depends on dertivatives of $b(x)$ only, which take on subplanckian values.}
\begin{align}\label{b0}
\frac{|\overline b(0)|}{M_{\rm Pl}}  \gg \, \sqrt{2\, \epsilon} \,  \mathcal N = \mathcal O(10), \quad \overline b(0) < 0~,
\end{align}
remains approximately constant until the end of inflation $t_{\rm end} \, H \simeq \mathcal N$,
with $\mathcal N$ the number of e-foldings, which phenomenologically can be taken to be  $\mathcal N \simeq 60-70$~\cite{inflation}. This contributes to the effective action an approximate de Sitter 
term~\cite{basilakos,ms2}
\begin{align}\label{lambda}
\mathcal S_\Lambda  &=
\sqrt{\frac{2}{3}}\,
\frac{\alpha^\prime}{96 \, \kappa} \, \int d^4 x \sqrt{-g} \, \langle \overline b \, R_{\mu\mu\rho\sigma}\, \widetilde R^{\mu\nu\rho\sigma} \rangle  \equiv  -  \int d^4x \, \sqrt{-g} \, \frac{\Lambda (H)}{\kappa^2} \nonumber \\ & \simeq   \int d^4 x \, \sqrt{-g}\, \Big(5.86 \times 10^{-5}  \, \Big(\frac{\mu}{M_s}\Big)^4 \, \sqrt{2\, \epsilon} \,
\Big[\frac{\overline b(0)}{M_{\rm Pl}}\Big] \, H^4 \Big)\,.
\end{align}
There are contributions to the total energy ($\rho_{\rm total}$) and pressure ($p_{\rm total}$) densities for this fluid coming from: the condensate \eqref{lambda} (superscript  ``condensate''), the KR axions (superscript  ``$b$'') and the gravitational-anomalies (Chern-Simons) fluctuations (superscript gCS): 
\begin{align}\label{stiffvacuumexcit}
\rho_{\rm total} =  \rho^b + \rho^{\rm gCS} + \rho^{\rm condensate} , \quad
p_{\rm total} = p^b + p^{\rm gCS} + p^{\rm condensate},
\end{align}
Using properties of the KR axion ``matter'' and the Cotton tensor, discussed in section \ref{sec:string}, we can explicitly 
demonstrate the following equations of state (for details we refer the reader to ref.~\cite{ms1}):
\begin{align}\label{bgCS}
p^b = + \rho^b, \, \, p^{\rm gCS} = \frac{1}{3} \, \rho^{\rm gCS} \,\, \Rightarrow \,\,
p^b + p^{\rm gCS} = \rho^b + \frac{1}{3} \, \rho^{\rm gCS} = -\frac{1}{3}\, \rho^{\rm gCS} = - (\rho^b + \rho^{\rm gCS})>0~,
\end{align}
and 
\begin{align}\label{dS}
p^{\rm condensate} = - \rho^{\rm condensate} \,< \,0.
\end{align}
From \eqref{bgCS} we observe that, were it not for the condensate \eqref{lambda}, the gravitational anomalies, due to their negative contributions to the energy density, would make that fluid
behave like ``phantom matter", with negative energy density and positive pressure, violating the weak energy condition~\cite{phantmat1,phantmat2} (which, by the way, is the kind of ``exotic matter" required for stabilisation of traversable wormholes~\cite{worm1,worm2}). Curiously, in our case, the equation of state of this part of the fluid is of RVM type~\cite{rvm,rvm2,rvm3,solahere}, \eqref{bgCS}. 
Nonetheless, the dominance of the condensate term \eqref{lambda} 
in the early Universe, which scales like $H^4$ and is characterised by the standard 
de Sitter equation of state \eqref{dS}, renders the total energy density {\it positive}~\cite{basilakos2,ms2}: 
\begin{align}\label{toten}
0 <  \rho_{\rm total}  \simeq  3\kappa^{-4} \, \Big[ -1.65 \times 10^{-3} \Big(\kappa\, H \Big)^2
+ \frac{\sqrt{2}}{3} \, |\overline b(0)| \, \kappa \, \times {5.86\, \times} \, 10^6 \, \left(\kappa\, H \right)^4 \Big]
\end{align}
under the condition \eqref{b0}. In view of \eqref{bgCS}, \eqref{dS}, the total equation of state is that of the conventional RVM~\cite{rvm,rvm2,rvm3,solahere}, a de Sitter type contribution, but with a mild cosmic-time dependence of the energy and pressure densities, due to their dependence on $H(t)$: 
\begin{align}\label{rvmeos}
p_{\rm total} (H(t)) = - \rho_{\rm total} (H(t)) \, <\, 0.
\end{align}
The energy density \eqref{toten} also has the form of a conventional RVM energy density~\cite{rvm,rvm2,rvm3,solahere}, 
\begin{equation}\label{rLRVM}
\rho^{\Lambda}_{\rm RVM}(H) = \frac{\Lambda(H)}{\kappa^2}=
\frac{3}{\kappa^2}\left(c_0 + \nu H^{2} + \alpha
\frac{H^{4}}{H_{I}^{2}} + \dots \right) \, >\, 0\;,
\end{equation}
with $H_I$ the inflationary scale inferred from Planck Collaboration data~\cite{Planck}, \eqref{PlHI}, and $c_0 \ge 0 $,  $\nu$ and $\alpha$ constants. In the conventional RVM, $\nu > 0$ and $\alpha >0$, 
however in our stringy-RVM inflationary phase, the coefficient 
of the $H^2$ term in \eqref{toten} is {\it negative}, due to the gravitational anomalous Chern-Simons contributions. 
There is also no evidence for the presence of a non-zero constant $c_0$ in this early RVM-inflationary phase, although such a (positive, cosmological) constant can be generated at late eras of the Universe evolution~\cite{basilakos2}, as we shall discuss in the next section \ref{sec:postinfl}.

So far, we have calculated the condensate \eqref{lambda} assuming an approximately  constant $H$. The consistency of the approach, and thus the emergence of an RVM inflation dynamically,  comes  about by considering the temporal evolution of the RVM  fluid \eqref{toten}, \eqref{rvmeos}, which, as we have just argued, arises as a consequence of the GW-induced  axion-gravitational-anomaly condensate \eqref{lambda}. Indeed, as discussed in \cite{rvmevol,rvmevol2}, on considering the conservation of the stress energy tensor of matter or radiation, with equation of state $\omega_m$, on a FLRW background space-time, in an RVM vacuum \eqref{rLRVM}, implies the evolution equation:
\begin{equation}\label{evol}
\dot H + \frac{3}{2} \, (1 + \omega_m) \, H^2 \, \Big( 1 - \nu - \frac{c_0}{H^2} - \alpha \, \frac{H^2}{H_I^2} \Big) =0~,
\end{equation}
which, on ignoring $c_0$ (which, as mentioned above, is consistent with our explicitly derived form \eqref{toten}), leads to a solution for $H(a)$ as a function of the scale factor $a$ (in units of the present-era scale factor) and the equation of state $\omega_m$ of  matter~\cite{rvmevol,rvmevol2}:
\begin{equation}\label{HS1}
 H(a)=\left(\frac{1-\nu}{\alpha}\right)^{1/2}\,\frac{H_{I}}{\sqrt{D\,a^{3(1-\nu)(1+\omega_m)}+1}}\,,
\end{equation}
where $D>0$ is an integration constant. In our stringy RVM case, the only ``matter" existing in the early stringy Universe is the stiff KR axion matter with $w_m=+1$, while the coefficient $\nu= -1.65 \times 10^{-3} < 0$. For the early Universe, $ a \ll 1$, and thus one may assume without loss of generality that
$D\,a^{3(1-\nu)(1+\omega_m)} \ll 1$. On account of \eqref{HS1}, then, this leads to
an (unstable) dynamical early de Sitter phase, characterised by
an approximately constant $H_{\rm de~Sitter}  \simeq \left(\frac{1-\nu}{\alpha}\right)^{1/2}\,\, H_{I}$.
Hence, the (positive) $H^4$ term in \eqref{toten} dominates  the early stages of the Universe evolution, and leads to inflation, as we consistently assumed when computing the condensate \eqref{lambda}. 
In our case, we can arrange for $\alpha =\mathcal O(1)$, and thus $H \simeq H_I$, as required for consistency, by adjusting appropriately the boundary value $|\overline b(0)| \kappa \sim 3.6 \times 10^3$ in agreement with the condition \eqref{b0}.  This RVM inflation is therefore dynamical, and does not involve external inflaton fields. Nonetheless, there is a {\it slow-roll} pseudoscalar field present, the KR axion \eqref{slowrollb}, but it does {\it not} itself cause the inflation.  

The early-Universe dominant dark energy component in \eqref{toten}, scaling as $H^4$, is purely geometric in origin, as it owes its existence to a primordial-GW-induced condensate of a CP-violating interaction term of the KR axion background field with the gravitational anomaly term, \eqref{lambda}. We note at this stage that, in the context of perturbative string theory~\cite{string}, formulated in given space-time backgrounds, like the FLRW Universe, local field redefinition ambiguities, which leave the perturbative scattering matrix invariant, according to the equivalence theorem of field theory~\cite{equiv1,equiv2}, allow for the quadratic-in-(gravitational-)curvature terms in the low-energy string effective action to be cast always in the (ghost-free) Gauss-Bonnet form~\cite{amb1,amb2,amb3,amb4}, which, in four space-time dimensions, is a total derivative. Hence, for constant dilatons we are considering here, such terms will not contribute to the action. Since an $H^4$ term is of the same order as terms quadratic in curvature, this leaves the CP-violating Chern-Simons anomalous gravitational terms as the sole contributor of $H^4$ to the vacuum energy density in our model through the aforementioned GW-induced-condensate mechanism.

We stress once more, that the condensate \eqref{rrt2} involves integration over graviton degrees of freedom, and, as such, it is a consequence of  (weak) quantum-gravity effects. 
We also remark~\cite{ms2} that the crucial for inflation $H^4$ term in the RVM energy density \eqref{rLRVM} 
is missing in {\it both}, the quantum field theory example of a non-minimally coupled scalar field considered in \cite{rvmqft} and the Starobinsky inflation~\cite{staro}. In fact, we remark for completeness, that the latter is based on $\dot H \simeq {\rm constant}$, rather than $H \simeq \rm constant$, which is the case of the RVM inflation~\cite{rvmevol2}. 

\section{Post-RVM-Inflationary eras \label{sec:postinfl}}

{\it Chiral fermionic} matter  and radiation (gauge fields) are assumed to be generated in our model at the end of the RVM inflationary epoch, as a result of the decay of the RVM vacuum~\cite{basilakos,basilakos2,ms1,ms2}. 
The corresponding effective action for chiral fermions is obtained in a similar manner as its bosonic counterpart  \eqref{sea4}. The torsion interpretation of the field strength $\mathcal H_{\mu\nu\rho}$ of the antisymmetric-tensor field, dictates the presence~\cite{basilakos2} of a linear interaction between the axial fermion current $J^{5 \mu}~\equiv~\sum_{i=\rm fermion~species} \, \overline \psi_i \gamma^5 \, \gamma^\mu \, \psi_i $ and 
$\epsilon_{\mu\nu\rho\sigma}\, \mathcal H^{\nu\rho\sigma} $, which, upon implementing the 
Bianchi identity constraint \eqref{modbianchi2} via the Lagrange multiplier KR axion-like field $b(x)$, and integrating out the $\mathcal H_{\mu\nu\rho}$ field strength in the path-integral, leads 
to the effective action with fermions~\cite{basilakos2}: 
\begin{align}\label{sea6}
S^{\rm eff} &=\; \int d^{4}x\sqrt{-g}\Big[ -\dfrac{1}{2\kappa^{2}}\, R + \frac{1}{2}\, \partial_\mu b \, \partial^\mu b -  \sqrt{\frac{2}{3}}\,
\frac{\alpha^\prime}{96\, \kappa} \, \partial_\mu b(x) \, {\mathcal K}^\mu
\Big] \nonumber \\
&+ S_{\rm Dirac~or~Majorana}^{Free} + \int d^{4}x\sqrt{-g}\, \Big[\Big( {\mathcal F}_\mu + \frac{\alpha^\prime}{2\, \kappa} \, \sqrt{\frac{3}{2}} \, \partial_{\mu}b \Big)\, J^{5\mu}    - \dfrac{3\alpha^{\prime\, 2}}{16 \, \kappa^2}\,J^{5}_{\mu}J^{5\mu}  + \dots \Big] + \dots,
\end{align}
where ${\mathcal F}^d  =   \varepsilon^{abcd} \, e_{b\lambda} \,  \partial_a \, e^\lambda_{\,\,c}$, with $e^\mu_{\,\,c}$ the vielbeins (with Latin indices pertaining to the tangent space of the space-time manifold at a given point, in a standard notation), 
$S_{\rm Dirac~or~Majorana}^{Free}$ denotes the free-fermion kinetic terms, 
and the $\dots$ in (\ref{sea6}) indicate gauge field kinetic terms, as well as terms of higher order in derivatives. The action \eqref{sea6} is valid for both Dirac or Majorana fermions (as would be the case of RHN fields that are also generated at the end of the stringy-RVM inflation). 
The reader is invited to take note of the presence in \eqref{sea6} of the CP-violating interactions of the derivative of the field $b$ with the axial fermion current $J^{5\mu} $, as well as of the repulsive axial-fermion-current-current term, 
$-\dfrac{3\alpha^{\prime\, 2}}{16 \, \kappa^2}\,J^{5}_{\mu}J^{5\mu}$, which is characteristic of theories with Einstein-Cartan torsion~\cite{torsion,torsion2shap}, as is our string-inspired model~\cite{kaloper}.  

\subsection{Cancellation of gravitational anomalies, generation of chiral anomalies and KR-axion mass \label{sec:kram}}

The chiral fermions are characterised by their own gravitational and chiral anomalies~\cite{alvarez}. The basic assumption of \cite{basilakos,basilakos2,ms1,ms2} was that the gravitational anomalies due to chiral fermions {\it cancel} the primordial gravitational anomalies due to the Green Schwarz counterterms \eqref{modbianchi2}. In this way, the standard cosmology for matter and radiation is more or less maintained, given that the corresponding quantum field theory is developed in a gravitational-anomaly-free background. 

However, in general, chiral anomalies, which involve (cosmic) gauge fields (electromagnetic or QCD gluon fields), remain~\cite{basilakos}: 
\begin{align}\label{anom2}
& \partial_\mu \Big[\sqrt{-g}\, \Big(  \sqrt{\frac{3}{8}} \frac{\alpha^\prime}{\kappa}\, J^{5\mu}  -  \frac{\alpha^\prime}{\kappa}\, \sqrt{\frac{2}{3}}\,
\frac{1}{96} \, {\mathcal K}^\mu  \Big) \Big]   =   \sqrt{\frac{3}{8}} \, \frac{\alpha^\prime}{\kappa}\, \Big(\frac{\alpha_{\rm EM}}{2\pi}  \, \sqrt{-g}\,  {F}^{\mu\nu}\,  \widetilde{F}_{\mu\nu} + \frac{\alpha_s}{8\pi}\, \sqrt{-g} \, G_{\mu\nu}^a \, \widetilde G^{a\mu\nu} \Big)~,
\end{align}
where $F_{\mu\nu}$ denotes the electromagnetic (EM) Maxwell tensor,  and $G_{\mu\nu}^a$ is the gluon field strength, with $a=1, \dots 8$ an adjoint SU(3) colour index, $\alpha_{\rm EM}$ the electromagnetic fine structure constant, and $\alpha_s$ the strong-interactions fine structure constant. The presence of gluon terms in the effective action may trigger the generation of a non-perturbative (through instantons) potential for the KR axion, during the QCD epoch of the Universe, during the post inflationary era. In this way a non-perturbative mass for the KR axion can be generated, as discussed in some detail in \cite{basilakos,basilakos2}. For the string mass scale \eqref{msscale}, it can be shown that the generated mass of the KR axion would be in the range  $1.17 \times 10^{-8} \lesssim m_b/(\rm eV) \lesssim 1.17 \times 10^{-5}$,
which is phenomenologically acceptable. In this way, the KR axion field, which is dual to the totally antisymmetric torsion \eqref{dual}, could play the r\^ole of a dominant component of DM, 
and hence, in conjunction with the aforementioned geometric interpretation of the dark energy sector of the stringy RVM Universe during inflation, one obtains a geometric interpretation of the entire dark sector (energy and matter) of this stringy Universe.

\subsection{KR axion-induced Lorentz- \& CPT-Violating Leptogenesis}

In addition, one also obtains a {\it geometric origin} of the {\it matter-antimatter asymmetry} in the Cosmos, since the presence of the undiluted LV KR axion backgrounds \eqref{slowrollb} at the end of inflation triggers leptogenesis during the early radiation epoch, in models with RHN~\cite{basilakos,basilakos2,ms2}, as we now come to discuss. To this end, we first note that, at the exit from inflation, the cancellation of gravitational anomalies implies the following equation of motion for the KR axion field $b(x)$:
\begin{align}\label{bEB}
\partial_{\alpha}\Big[\sqrt{-g}\, \partial^{\alpha}\bar{b} \Big] = \rm chiral~anomalies~(when~dominant),
\end{align}
In \cite{basilakos2} we assumed that the chiral anomaly terms become important at late stages of the radiation era. In the absence of chiral anomalies, which is the case of the epochs after the exit from RVM inflation in our string-inspired cosmology, one obtains from \eqref{bEB} (upon setting the right-hand-side to sero) that $\dot b \propto a^{-3}(t) \sim T^3 $, where $T$ is the temperature of radiation in the early Universe (assuming the scaling $a(t) \sim T^{-1}$). Matching the value \eqref{slowrollb} at the exit from the RVM inflation, where the temperature is taken to be of order of the 
Gibbons-Hawking temperature~\cite{GH} of the de Sitter space time,  $T_{\rm GH} = H_I/(2\pi)$, with the solution of \eqref{bEB} (with zero right-hand side) at a temperature $T$, one obtains
for the early stages of radiation~\cite{basilakos2}:
\begin{align}\label{bscale2}
\dot{b} \simeq  3.5 \times 10^{11} \, M_{\rm Pl}^2 \, \Big(\frac{T}{M_{\rm Pl}}\Big)^3 .
\end{align}

As discussed in \cite{sarkar3,sarkar4,sarkar6}, and we shall describe briefly below, 
such backgrounds can produce phenomenologically correct leptogenesis in theories with RHN in their spectra. In particular, we consider lepton-number asymmetry originating from tree-level decays of heavy sterile RHN  into SM leptons. The relevant part of the effective Lagrangian is given by:
\begin{align}\label{smelag}
\mathcal{L}= {\mathcal L}_{\rm SM} + i\overline{N}\, \gamma^\mu\, \partial_\mu \, N-\frac{m_N}{2}(\overline{N^{c}}N+\overline{N}N^{c})-\overline{N}\gamma^\mu\, B_\mu \, \gamma^{5}N-\sum_f \, y_{f}\overline{L}_{f}\tilde{\phi}^dN+ {\rm h.c.}
\end{align}
where h.c.  denotes hermitian conjugate, ${\mathcal L}_{\rm SM}$ denotes the SM Lagrangian,
$N$ is the RHN field, of (Majorana) mass $m_N$,  $\tilde \phi$ is the SU(2) adjoint of the Higgs field  $\phi$ ($\tilde{\phi}^d_i \equiv \varepsilon_{ij}\phi_j~, \, i,j=1,2,$ SU(2) indices),
 and $L_{f}$ is a lepton (doublet) field of the SM sector, with $f$ a generation index, $f=e, \mu, \tau$, in a standard notation for the three SM generations; $y_f$ is a Yukawa coupling, which is non-zero and provides a non-trivial (``Higgs portal'') interaction between the RHN and the SM sector, used in the seesaw mechanism for generation of SM neutrino masses. The  quantity $B_\mu$   is defined as
\begin{align}\label{background}
B_\mu = M_{\rm Pl}^{-1} \, \dot{\overline b}\, \delta_{\mu0}\,,
\end{align}
and denotes the LV,  CPV and CPTV background \eqref{bscale2}, with  $B_\mu$ having only a temporal component. For the short period of leptogenesis, such a background is viewed
as slowly varying in the cosmic frame~\cite{sarkar6}, and the Lagrangian (\ref{smelag}) assumes approximately the form of a Standard Model Extension (SME) Lagrangian in a LV and CPTV (constant) background~\cite{sme}.

In the context of the model \eqref{smelag}, a lepton asymmetry is then generated due to the CPV and CPTV tree-level decays of the RHN $N$ into SM leptons in the presence of the background \eqref{background}~\cite{sarkar3,sarkar4,sarkar6}: ${\rm Channel} ~I: \,  N \rightarrow l^{-}h^{+}~, ~ \nu \, h^{0}$,  and 
${\rm Channel ~II}: \, N \rightarrow l^{+}h^{-}~,~  \overline \nu \, h^{0}$,
where $\ell^\pm$ are charged leptons, $\nu$ ($\overline \nu$) are light, ``active", neutrinos (antineutrinos) in the SM sector, $h^0$ is the neutral Higgs field, and
 $h^\pm$ are the charged Higgs fields, which, at high temperatures, above the spontaneous electroweak symmetry breaking, of interest in this scenario, do not decouple from the physical spectrum.  As a result of the non-trivial $B_0 \ne 0$ background (\ref{background}), \eqref{bscale2}, the decay rates of the Majorana RHN between the channels I and II are different, resulting in a Lepton asymmetry (we assume for brevity and concreteness that the Lepton asymmetry is produced mainly by the decay of one family of RHN, of mass $m_N$ - extension to three generations is straightforward)~\cite{sarkar6},
 \begin{align}\label{lepto}
 \frac{ \Delta L^{TOT}(T=T_D)}{s} \sim  q\,  \dfrac{B_{0}(T_D)\, m_N^2}{T_D^3} \sim \, q\, 3.5 \times 10^{11} \,
 \Big(\dfrac{m_N}{m_{\rm Pl}}\Big)^2, \quad q > 0,
 \end{align}
 where we used  \eqref{bscale2}, \eqref{background}, $s$ is the entropy density of the Universe,  $T_D \simeq m_N$ denotes the freeze-out temperature, and $0 < q=\mathcal O(10)$ is a numerical factor expressing theoretical uncertainties in the approximate analytic methods used for the evaluation of \eqref{lepto}~\cite{sarkar6}. The lepton asymmetry \eqref{lepto}  can then be communicated to the baryon sector via Baryon-minus-Lepton-number ($B-L$) conserving sphaleron processes in the SM sector~\cite{smbau}. The observed amount of baryon asymmetry (baryogenesis)  in the Universe \eqref{bau}~\cite{Planck} is then obtained by requiring that
the lepton asymmetry \eqref{lepto} is of the same order as \eqref{bau}, which implies
$T_D \sim m_N \sim  \sim 10^7$ GeV, compatible with seesaw mechanisms and  Higgs-mass stability criteria~\cite{sarkar3,sarkar4,sarkar6,basilakos2}.

\section{Modern-era Phenomenology of stringy RVM \label{sec:modern}}

After leptogenesis, during the radiation era, the temperature of the Universe continues to drop at a rate $a(t) \sim 1/T$, until the expansion of the Universe is such that the terms due to the chiral anomalies in the solution for the KR axion background  \eqref{bEB}, which scale as $a^{-2}(t)$~\cite{basilakos2}, dominate over the $a^{-3}(t)$-scaling terms. The detailed phenomenological analysis of \cite{basilakos2}, then, demonstrates that, 
in the modern era, under the assumption of having an amplitude $B(t_0)$ for the cosmic magnetic field intensity today (with $t_0$ the age of the (observable) Universe), contributing to the chiral electromagnetic anomaly, one obtains for the matter-dominated and present eras:
\begin{align}\label{bsolmatt2}
\dot{b}\,\Big|_{{\rm matter~era}} &\simeq \Big(\frac{M_{\rm Pl}}{M_s}\Big)^2 \, \sqrt{\frac{3}{2}} \, \frac{e^2}{4\pi^2} \, \frac{{B}^2(t_0)}{k \, M_{\rm Pl}^3} \, T^2,
\end{align}
where we assumed for concreteness monochromatic configurations for the magnetic field, corresponding to a single mode of momentum $k$. In our stringy-RVM model $M_{\rm Pl}/M_s$ is determined by \eqref{msscale}. 
This $T^2$ scaling for the $\dot{b}$ KR axial background during the late eras of the Universe  prompts one to conjecture that today 
 \begin{align}\label{krbtoday}
 \dot{\overline b}_{\rm today} \sim \sqrt{2\epsilon^\prime}\, H_0 \, M_{\rm Pl},
 \end{align}
 in analogy to \eqref{slowrollb}, with $H_0$ the present-epoch Hubble parameter. From \eqref{bsolmatt2}, this allows for an estimate of $\epsilon^\prime$ as a function of the amplitude of the present-era cosmic magnetic field intensity $B(t_0)$~\cite{basilakos2}. The latter is a phenomenological parameter, and cannot be determined from first principles in string theory, in view of the ground-state string landscape issue. Nonetheless, if we take into account that in the presence of a $B(t_0) \ne 0$, the cosmic energy density acquires a component $\rho_0^B=\frac{1}{2}\, B^2(t_0)$, 
 we observe~\cite{basilakos2} that $\epsilon^\prime \simeq \frac{B(t_0)^2}{M_{\rm Pl}^2 \, H_0^2} = \frac{2}{3} \frac{\rho_0^B}{\rho_c^{(0)}}$, with $\rho_c^{(0)} = H_0^2 \, M_{\rm Pl}^2/3$ the critical density of the universe. On using the fact~\cite{basilakos2} that the slow-roll parameter of $b(x)$ measures the ratio of its kinetic energy, $K_b\sim (1/2)\,\dot{b}^2$,  to the critical energy density of the Universe, 
$\rho_c=(M_{\rm Pl} H)^2/3$, and making the further phenomenological assumption that this KR axion field plays the r\^ole of the dominant DM component in the Universe~\cite{basilakos,basilakos2}, with $K_b \sim 0.1\, U$, where $U$ is its potential energy (as typical in slow-roll, quintessence-like, field situations), 
one can get the DM content in the right ballpark~\cite{Planck}:
$\Omega_{m0}=\frac{\rho_{m0}}{\rho^{(0)}_c}\simeq \frac{U}{\rho^{(0)}_c}\simeq 10\, \frac{K_b}{\rho^{(0)}_c}\simeq 10\, \epsilon={\cal O}(0.1)\,,$
with $\rho_{m0}$  the current energy density of DM in the universe, provided 
 \begin{align}\label{eeprime}
 \epsilon \sim \epsilon^\prime ={\mathcal O}(10^{-2}).
 \end{align}
where $\epsilon$ defines the slow roll of $\dot b$ during the RVM inflationary era \eqref{slowrollb}. 
This ``coincidence issue" needs to be understood better, in terms of microscopic string theory models.
For our phenomenological purposes here, we only note that the result \eqref{eeprime}, implies a KR axion background field \eqref{background} in the current era of order 
$B_0 \Big|_{\rm today} \sim 2.435 \times 10^{-34}\, {\rm eV}$,
which is comfortably compatible with the upper bounds of LV and CPTV torsion today~\cite{smebounds}, $B_0 < 10^{-2} \,{\rm  eV}$ for the temporal component,  and  $B_i < 10^{-31}$~GeV, $i=1,2,3$ for the spatial components (in our case, the latter are derived from the temporal component \eqref{background} in the cosmic frame, by applying a Lorentz boost, due to our relative velocity $\vec v$ with respect to that frame, $B_i = \gamma \frac{v_i}{c}\, B_0,\quad i=1,2,3$, which is measured as $|\vec v| ={\mathcal O}(390 \pm 60)$~Km/sec~\cite{Planck}). 

In the current era, the cosmic energy density of this string-inspired cosmology also assumes an RVM form~\cite{basilakos,basilakos2}, 
\begin{align}\label{modernDE}
\rho_0 = \frac{3}{\kappa^2} \Big(c_0 + \nu_0 \, H_0^2 \Big)~, 
\end{align}
where now $\nu_0 > 0$, due to contributions from cosmic electromagnetic background fields~\cite{basilakos2} (terms of order $H_0^4$ in the cosmic energy density are negligible in the current epoch). Comparison with the data~\cite{rvmpheno1,rvmpheno,rvmpheno2,tsiapi,solahere} indicates
$\nu_0 = \mathcal O (10^{-3})$. The presence of a cosmological constant, $c_0$ in \eqref{modernDE}, required by the current-era phenomenology~\cite{rvmpheno1,rvmpheno}, is still not understood microscopically in our string-inspired  cosmological model.
One potential explanation is that $c_0$ is associated with a Lorentz-invariant condensate of 
the (torsion-induced) last term on the right-hand-side of \eqref{sea6},
which leads to a constant vacuum energy density~\cite{popl,popl2}  $[3\alpha^{\prime\, 2}/(16 \, \kappa^2)]\,  \langle J^{5}_{\mu}J^{5\mu}  \rangle$.
The condensate may be created after leptogenesis in our model, at late eras of the Universe, when the latter has sufficiently cooled down. If this scenario is realised in nature, it would imply that the current-epoch dark energy, responsible for the acceleration of the Universe, has also a {\it geometrical origin}, in view of the association of the condensate with torsion in our stringy model.  

Other scenarios accounting for the presence of a $c_0$ term in \eqref{modernDE} could arise in embeddings of the stringy RVM model in brane cosmologies. Indeed, in such frameworks, the 
cosmological constant could 
be due to contributions of bulk higher-curvature terms on our brane Universe, in warped higher-dimensional geometries~\cite{rizos}. 

Regarding the alleviation of tensions, provided, of course, that the latter persist in future data analyses, 
we remark that the so-called type II RVM is the best for alleviating 
simultaneously the $H_0$ and $\sigma_8$ tensions~\cite{rvmpheno2}. Such a variant of RVM contains, in  its energy density, additional terms, amounting to a mild cosmic time dependence of an effective Newton constant G$_{\rm eff}(H) = G/\varphi (t)$, where $\varphi(t)$ is a phenomenological auxiliary parameter to parametrise the mild cosmic time dependence. 

In \cite{ms1} we speculated that we could obtain somewhat analogous modifications in our stringy RVM. Indeed, as we can infer from our supergravity model~\cite{brokensugra}, which is embeddable into our stringy framework, such corrections could be obtained as a result of integrating out graviton fluctuations, about a de Sitter background, which also approximately characterises the current era of the Universe~\cite{Planck}. 
Such corrections depend on the one-loop-induced~\cite{tseytlin,brokensugra} positive de Sitter parameter, which is identified with the cosmological constant $\Lambda \sim 3H^2$, or, equivalently, with the (background) curvature scalar,
$\widehat R \sim 12 H^2$, in exactly de Sitter space-times, so general covariance is respected.\footnote{In the absence of any mass parameter, as appropriate for vacuum solutions, one may pass from a local de Sitter space-time to a global (cosmological one) by means of appropriate general coordinate transformations~\cite{lanczos}, which leave the effective action invariant. It is in this sense that we can link the results of \cite{tseytlin,brokensugra} about a local de Sitter space-time background to our cosmological case.}
The relevant correction  terms in the one-loop effective Lagrangian density have the structure~\cite{ms1} 
\begin{equation}\label{1looplagr}
\delta \mathcal L^{\rm 1-loop}_{\rm quant.~grav.} = \sqrt{-\widehat g}\, \Big[ \widetilde \alpha_0 + \widehat R (c_1 + c_2\, {\rm ln}(\frac{1}{12}\kappa^2 \widehat R)) \Big] + \dots 
\end{equation}
with $c_i \propto \kappa^2 \mathcal E_0, i=1,2,$ appropriate constants and $\mathcal E_0$ a bare (constant) vacuum energy density scale (the ellipses $\dots $ denote terms of higher order in $\widehat R = 12 H^2$, which are not dominant in the current epoch $(H = H_0)$). These structures seem pretty generic for gravitational models involving weak quantum gravity corrections about such backgrounds~\cite{tseytlin}. 

 On using the graviton equations for the one-loop corrected effective Lagrangian, it can be shown that the terms \eqref{1looplagr} lead to corrections to the effective stress-energy tensor in the current era, 
\begin{equation}\label{1loopenden}
\delta \rho_0^{\rm vac}  =  \frac{1}{2}\widetilde \alpha_0 + 3 (c_1-c_2) H_0^2 + 3 c_2 H_0^2 \, {\rm ln}(\kappa^2 H_0^2)~, 
\end{equation}
which should be added to \eqref{modernDE}.\footnote{Such one-loop quantum-gravity corrections, and also those of the form $H^4{\rm ln}(\kappa^2 H^2)$\cite{brokensugra,ms1}, which should be taken into account for the early eras of the Universe, are subdominant compared to the anomaly-induced $H^4$ term in \eqref{toten}, for $\kappa^4 |\mathcal E_0| < 1$, and thus our conclusions on RVM inflation are not affected.} The supergravity example~\cite{brokensugra} indicates that the one-loop correction term $\frac{1}{2} \widetilde \alpha_0$ is constant, independent of ${\rm ln} H^2$ terms. 
This will lead to differences from the standard type II RVM~\cite{rvmpheno2}. We hope to study these issues, and their phenomenology, in detail in a future work.

\section{Conclusions and Outlook \label{sec:concl}}

In this work, I reviewed a string-inspired cosmology with gravitational anomalies and a KR axion, stemming from the massless gravitational multiplet of the string. 
I have argued in favour of a geometric interpretation of {\it both} the Dark Energy and Dark Matter sectors of this stringy Universe.
At early stages, the dark energy sector of this universe assumes an RVM form, with the dominant terms
scaling like $H^4$, with $H$ the corresponding Hubble parameter, which are due to condensates of the gravitational anomalies due to primordial GW. Such terms lead to inflation of RVM type, without external inflatons. 
The GW may originate in a pre-RVM-inflationary epoch. During this inflation, there is an undiluted KR axion background, which violates spontaneously Lorentz and CPT symmetries (LV \& CPTV), and survives the post inflationary era, leading to matter-antimatter asymmetries during the radiation epoch. The KR axion can acquire a non trivial mass through QCD instanton effects, during the QCD epoch, 
thus playing the r\^ole of a Dark Matter component.

There are several aspects of this work that we would like to pursue further. First, we would like to see if the presence of the LV and CPTV KR axion background leaves detectable imprints in the CMB spectrum. Second, given the fact that the coefficient of the $H^2$ term in the RVM-like energy density during inflation is negative, due to the gravitational anomalies, it would be interesting to see if one could somehow find evidence for it in the early Universe data. This is not an easy task, given that the exit from inflation in our model involves a phase transition. Another phenomenologically interesting avenue for research is the incorporation of the other axions that arise after compactification in phenomenologically realistic string models, which lead to a rich phenomenology~\cite{axiverse,marsh}. Big Bang nucleosynthesis constraints should also be carefully taken into account in the post-inflationary era of the model. We also mention that, so far, there is no observational evidence on the existence of Chern-Simons gravity in the late Universe~\cite{yunrecent}, which is in agreement with the basic feature of our proposed stringy RVM cosmology on the cancellation of gravitational anomalies in the post-inflationary epochs.
Phenomenological comparison of our string-inspired cosmological model, which is characterised by a totally antisymmetric torsion, with other cosmologies with torsion (e.g. \cite{tors1,tors2}), as well as teleparallel-gravity models~\cite{tele}, is another interesting avenue for future research. I stress once again, though, that torsion in our stringy RVM is totally antisymmetric, associated with the string-model independent KR axion field. In this respect, our model is different from generic torsion models, where the torsion has other components. Moreover, our stringy cosmology has necessarily graviton fields in its physical excitation spectrum, and thus our stringy RVM is different in many respects from 
teleparallel-gravity models. Last, but not least, we mention the need to understand, in a quantitative manner, quantum-gravity corrections depending logarithmically on $H^2$, and study their phenomenology associated with the 
current-era tensions in the data, comparing the model with the type II RVM.

I hope that the above discussion made it clear to the reader that, in order to constrain potential new models for the dark sector of the Universe, including its inflationary epoch, one needs to combine multi-messenger observations, not only cosmic but also terrestrial, for instance precision measurements that currently bound Lorentz-violating (LV) effects of torsion in the laboratory~\cite{smebounds}. 
In our stringy RVM cosmology, spontaneous LV and torsion play crucial r\^oles in several aspects. I conclude by remarking that {\it LV} might well be a generic feature of {\it quantum gravity}~\cite{qglv}. However, in this review, I also claimed that {\it primordial gravitational anomalies} may be another feature of quantum gravity models, which could provide a reason for our existence, namely the dominance of matter over antimatter in the Cosmos, via spontaneous CP and CPT violating leptogenesis. In this sense, both, the dark sector of the Universe and our existence, appear to have geometrical origins. If imprints of such cosmic LV and CPTV signatures are discovered in, say, data from early Universe cosmology ({\it e.g.} CMB spectra), then, depending on their magnitude, one might obtain crucial tests for the model and, in general, for other models of LV quantum gravity that exist in the current scientific literature. {\it Affaire \`a suivre ...}

\acknowledgements

NEM thanks Prof. S. Cotsakis for the invitation to contribute to this Special Issue of the Roy. Soc. (London) Phil. Trans  A on {\it The future of mathematical cosmology}. He also acknowledges participation in the COST Association Action CA18108 ``{\it Quantum Gravity Phenomenology in the Multimessenger Approach (QG-MM)}''. This work is funded in part by the UK Science and Technology Facilities  research Council (STFC) under the research grant ST/T000759/1.
This work does not use any data.
The author declares that he has no competing interests.


\end{document}